\documentclass[aps,nofootinbib,showpacs]{revtex4}
\usepackage{graphicx}

\begin{document}

\preprint{hep-th/0201188}

\title{Stability of the Anisotropic Brane Cosmology}

\author{Chiang-Mei Chen}
 \email{cmchen@phys.ntu.edu.tw}
 \affiliation{Department of Physics, National Taiwan University,
Taipei 106, Taiwan}

\author{W. F. Kao}
 \email{wfgore@cc.nctu.edu.tw}
 \affiliation{Institute of Physics, Chiao Tung University, Hsin Chu,
Taiwan}

\date{\today}

\begin{abstract}
The stability of the Bianchi type I anisotropic brane cosmology is
analyzed in this paper. We also study the effect of the brane
solution by comparing the models on the 3-brane and the models in
the conventional Einstein's space. Analysis is presented for two
different models: one with a perfect fluid and the other one with
a dilaton field. It is shown that the anisotropic expansion is
smeared out dynamically for both theories in the large time limit
independent of the models with different types of matter. The
initial states are, however, dramatically different. A primordial
anisotropic expansion will grow for the conventional Einstein's
theory. On the other hand, it is shown that the initial state is
highly isotropic for the brane universe except for a very
particular case. Moreover, it is also shown that the Bianchi type
I anisotropic cosmology is stable against any anisotropic
perturbation for both theories in the large time limit.
\end{abstract}


\maketitle

\section{Introduction}
The observation of the cosmic microwave background (CMB) radiation
\cite{data,cobe} indicates that our Universe is globally isotropic
to a very high degree of precision. Therefore, our Universe is
usually assumed to be described by the Friedmann-Robertson-Walker
(FRW) metric in most of the literatures. The origin of the
isotropic universe has also become an interesting research topic
ever since.

On the other hand, it is known that there is a small large-angle
CMB anisotropies, $\Delta T/T\simeq 10^{-5}$, under the CMB
background \cite{data,cobe}. The isotropy of our Universe may have
to do with the choice of initial conditions and the stability of
the evolutionary solutions. Therefore, the present isotropic phase
could be a dynamic result of the evolution of our Universe, no
matter what the initial state started out. In this paper, we will
address on this issue in both the standard Einstein's theory and
the newly developed brane world scenario \cite{RS99a,RS99b}. In
particular, we will consider the evolution of an anisotropic
cosmology described by the Bianchi type I models
\cite{m1}-\cite{barrow} for two different types of matter sources:
one is the perfect fluid and the other one is a dilaton field. Our
result shows that the anisotropic models considered all evolve
dynamically from the anisotropic Bianchi type I universe into the
isotropic FRW space in the large time limit. This property reveals
that the isotropy of the cosmological principle may be justified
and made consistent with our current observational data.

These two different theories give, however, completely different
initial anisotropic expansion at the very early stage of
evolution. Indeed, for the conventional Einstein's theory (CET),
the anisotropic expansion tends to be large in the very early
stage. In another words, the universe has to begin from a highly
anisotropic initial expansion and then decays to zero as the time
increases. On the other hand, in the brane world scenario
\cite{SMS00}-\cite{LMSW01}, due to the quadratic correction which
significantly changes the early time behavior of the Universe. As
a result, any initially non-vanishing anisotropy parameter tends
to vanish in the very early period. There is a characteristic
time, $t_c$, that divides the evolution of mean anisotropy
parameter into two different stages. The mean anisotropy parameter
is increasing when $t<t_c$ and reaches its maximal value at
$t=t_c$. After that, mean anisotropy parameter starts to decay.
This result remains true for both the model with a perfect fluid
and the model with a scalar field. And this appears to be a
general feature independent of the types of matter considered.

In addition, the stability analysis \cite{CHM02}-\cite{kao00}
indicates that all models in two different theories are all stable
against any anisotropic perturbation. In particular, we also show
that the system is stable against the dilaton perturbation when
the dilaton potential is close to its local minimum. Note that the
exact solution is known only when the dilaton potential is a
constant. This also justifies the stability of the isotropic
universe in the large time limit under the effect of a dilaton
potential with local minima.

This paper is organized as follows. In Sec. II we briefly review
the brane world formulation and the Bianchi type I model. Then we
consider the evolution of the anisotropic cosmology and its
stability in both theories of the CET and the brane world.
Analysis is presented for model coupled to a perfect fluid in
section III. The model with a dilaton field and a constant
potential is presented in section IV. Finally, the conclusion and
possible implications are drawn in Sec. V.

\section{Preliminary of Brane World and Anisotropic Cosmology}
The brane world scenario assumes that our Universe is a
four-dimensional space-time, a 3-brane, embedded in the 5D bulk
space-time. All the matter fields and the gauge fields except the
graviton are confined on the the 3-brane as a prior requirement
in order to avoid any violations with the empirical results.
Moreover, inspired by the string theory/M-theory, the
$Z_2$-symmetry with respect to the brane is imposed \cite{HW96}.
A formal realization of the brane world scenario, which recovers
the Newton gravity in the linear theory, is the Rundall-Sundrum
model \cite{RS99a,RS99b} in which the 4D flat brane(s) is
embedded in the 5D anti-de Sitter (AdS) space-time. Later on, a
covariant formulation of the effective gravitational field
equations on the 3-brane has been obtained via a geometric
approach by Shiromizu, Maeda and Sasaki \cite{SMS00,SSM00}. It is
shown that the effective four-dimensional gravitational field
equations on the brane take the the following form
\begin{equation} \label{BWEq}
G_{\mu\nu} = - \Lambda g_{\mu\nu} + k_4^2 T_{\mu\nu} + k_5^4
S_{\mu\nu} - E_{\mu\nu},
\end{equation}
where $G_{\mu\nu}$ and $T_{\mu\nu}$ are the Einstein and
energy-momentum tensors. $S_{\mu\nu}$ is a quadratic contribution
of $T_{\mu\nu}$ defined as
\begin{equation}
S_{\mu\nu} = \frac1{12} T T_{\mu\nu} - \frac14 T_\mu{}^\alpha
T_{\nu\alpha} + \frac1{24} g_{\mu\nu} \left(
3T^{\alpha\beta}T_{\alpha\beta}-T^2 \right).
\end{equation}
The effective 4D parameters, e.g., the cosmological constant
$\Lambda$ and gravitational coupling $k_4$, are determined by the
5D bulk cosmological constant $\Lambda_5$, the 5D gravitational
coupling $k_5$ and the tension of the brane $\lambda$ via the
following relations
\begin{equation}
\Lambda = k_5^2 \left( \frac{\Lambda_5}2 + \frac{k_5^2
\lambda^2}{12} \right), \qquad k_4^2 = \frac{k_5^4 \lambda}6.
\end{equation}
Here $g_{\mu \nu}$ is the metric tensor on the brane. In addition,
the quantity $E_{\mu\nu}$ is a pure bulk effect defined by the
bulk Weyl tensor \cite{SMS00}.

>From the Eq. (\ref{BWEq}), it is easy to realize that the brane
world scenario is different from the CET by two parts: (a) the
matter fields contribute local ``quadratic'' energy-momentum
correction via the tensor $S_{\mu\nu}$, and (b) there are
``nonlocal'' effects from the free field in the bulk, transmitted
via the projection of the bulk Weyl tensor $E_{\mu\nu}$.
Therefore, the CET can be treated as a limit of the brane theory
by taking $E_{\mu\nu}=0$ and $k_5\to 0$ with properly adjusted
values of the constants $k_4$ and $\Lambda$.

In addition to the generalized Einstein equations (\ref{BWEq}),
the energy-momentum tensor also satisfies the conservation law
$\nabla_\mu T^{\mu\nu} = 0$. Therefore, there is a constraint on
the tensor $E_{\mu\nu}$, $\nabla_\mu E^{\mu\nu} = k_5^4 \nabla_\mu
S^{\mu\nu}$, due to the Bianchi identity on the brane. Here, the
operator $\nabla$ is the covariant derivative with respect to the
metric $g_{\mu\nu}$ on the brane. One should point out here that
the field equations on the brane, namely the generalized Einstein
equations, the conservation of energy-momentum and the constraint
on $E_{\mu\nu}$ are, in general, not a closed system in the 4D
brane since the quantity $E_{\mu\nu}$ is five-dimensional. It can
only be evaluated by solving the field equations in the bulk. In
this paper, we will, however, only consider the quadratic effect
on the brane world in the anisotropic background. Therefore, we
will set $E_{\mu\nu}=0$ which is equivalent to embedding the
3-brane in the pure AdS bulk space-time.

Another important subject in this paper is the anisotropic
cosmology described by Bianchi type I metric. The line element of
the Bianchi type I space, an anisotropic generalization of the
flat FRW geometry, can be written as
\begin{equation}
ds^2 = -dt^2 + a_1^2(t) dx^2 + a_2^2(t) dy^2 + a_3^2(t) dz^2,
\end{equation}
with $a_i(t),\, i=1,2,3$ the expansion factors on each different
spatial directions. For later convenience, we will introduce the
following variables
\begin{eqnarray}
V &\equiv& \prod_{i=1}^3 a_i, \qquad\qquad \text{volume scale
factor}, \label{V} \\
H_i &\equiv& \frac{\dot a_i}{a_i}, \quad i=1,2,3, \qquad
\text{directional Hubble factors}, \\
H &\equiv& \frac13 \sum_{i=1}^3 H_i = \frac{\dot V}{3V}, \qquad
\text{mean Hubble factor}. \label{H}
\end{eqnarray}
In addition, we will also introduce two basic physical
observational quantities in cosmology:
\begin{eqnarray}
A &\equiv& \sum_{i=1}^3 \frac{(H_i-H)^2}{3H^2}, \qquad \text{mean
anisotropy parameter}, \\
q &\equiv& \frac{d}{d t} (H^{-1}) - 1, \qquad \text{deceleration
parameter}.
\end{eqnarray}
Note that $A \equiv 0$ for an isotropic expansion. Moreover, the
sign of the deceleration parameter indicates how the Universe
expands. Indeed, a positive sign corresponds to ``standard''
decelerating models whereas a negative sign indicates an
accelerating expansion.

\section{Anisotropic Universe with A Perfect Fluid}
In this section, we will consider the case that the matter
energy-momentum tensor, $T_{\mu\nu}$, is a perfect fluid whose
components are given by
\begin{equation}
T^0{}_0 = -\rho, \qquad T^1{}_1 = T^2{}_2 = T^3{}_3 = p .
\end{equation}
Here the energy density $\rho$ and the pressure $p$ of the
cosmological fluid obey a linear barotropic equation of state of
the form $p=(\gamma-1) \rho$ with $\gamma$ a constant in the range
$1 \leq \gamma \leq 2$.

\subsection{Conventional Einstein's Theory}

For the CET, the dynamics of the space-time is determined by the
Einstein equations and the energy-momentum conservation law
\begin{equation}
G_{\mu\nu} = - \Lambda g_{\mu\nu} + k_4^2 T_{\mu\nu}, \qquad
\nabla_\mu T^{\mu\nu} = 0,
\end{equation}
with $G_{\mu\nu}$ the Einstein tensor, $\Lambda$ the cosmological
constant and $k_4$ the gravitational coupling $k_4^2=8\pi G$. For
the Bianchi type I cosmology with a perfect fluid the $tt$- and
$ii$-components of Einstein equations
\footnote{ Actually, the equations (\ref{GRH}) and (\ref{GRHi})
are the time and spatial components of the following version of
the Einstein equation: $R_{\mu\nu}=\Lambda g_{\mu\nu} + k_4^2
(T_{\mu\nu}-\frac12 T g_{\mu\nu})$, with $R_{\mu\nu}$ the Ricci
tensor and $T$ the trace of $T_{\mu\nu}$.}
and the conservation law become
\begin{eqnarray}
3 \dot H + \sum_{i=1}^3 H_i^2 &=& \Lambda - \frac{3\gamma-2}2 \,
k_4^2 \rho, \label{GRH} \\
\frac1{V} \frac{d}{d t} (V H_i) &=& \Lambda - \frac{\gamma-2}2 \,
k_4^2 \rho, \quad i=1,2,3, \label{GRHi} \\
\dot \rho + 3 \gamma H \rho &=& 0. \label{GRrho}
\end{eqnarray}

\begin{figure}
\includegraphics[width=8cm]{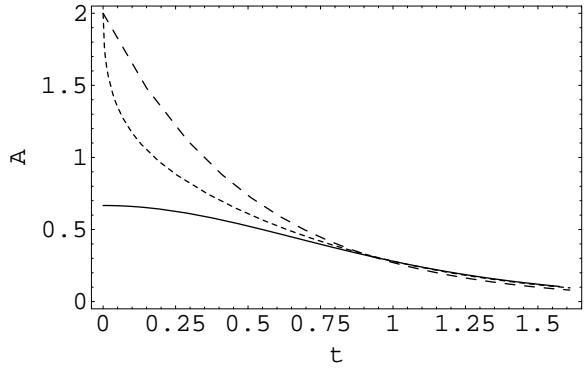}
\caption{\label{fig:F1} Mean anisotropy parameter of the Bianchi
type I conventional Einstein's theory with a perfect fluid:
$\gamma=2$ (solid curve), $\gamma=1.5$ (dotted curve) and
$\gamma=1$ (dashed curve). The normalization of the parameters is
chosen as $3\Lambda=1, 3k_4^2\rho_0=2$, and $C=1$.}
\end{figure}

\begin{figure}
\includegraphics[width=8cm]{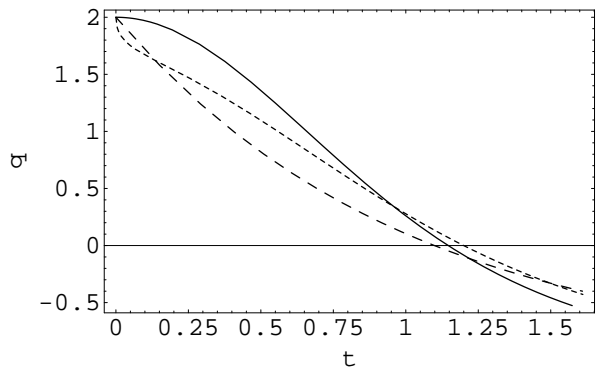}
\caption{\label{fig:F2} Deceleration parameter of the Bianchi type
I conventional Einstein's theory with a perfect fluid: $\gamma=2$
(solid curve), $\gamma=1.5$ (dotted curve) and $\gamma=1$ (dashed
curve). The normalization of the parameters is chosen as
$3\Lambda=1, 3k_4^2\rho_0=2$, and $C=1$.}
\end{figure}

First of all, the Eq. (\ref{GRrho}) can be easily solved to obtain
the time evolution law of the energy density of the fluid
\begin{equation}\label{rho}
\rho = \rho_0 V^{-\gamma}, \qquad \rho_0 = \text{constant} > 0.
\end{equation}
Next, by summing the Eqs. (\ref{GRHi}) with respect to the index
$i$, we can obtain the relation
\begin{equation}\label{dVH}
\frac1{V} \frac{d}{dt} (VH) = \Lambda - \frac{\gamma-2}2 \, k_4^2
\rho .
\end{equation}
Subtracting this result with the Eqs. (\ref{GRHi}), we can show
that
\begin{equation}\label{HiH}
H_i = H + \frac{K_i}{V}, \qquad i=1,2,3, \label{Hi}
\end{equation}
with $K_i, \, i=1,2,3$ the constants of integration satisfying the
consistency condition $\sum_{i=1}^3 K_i=0$. Moreover, by using the
evolution law of the matter energy density (\ref{rho}), the basic
equation (\ref{dVH}) describing the dynamics of the anisotropic
Universe can be written as
\begin{equation}
\ddot V = 3 \Lambda V - \frac{3(\gamma-2)}2 \, k_4^2 \rho_0
V^{1-\gamma} ,
\end{equation}
with the general solution
\begin{equation}\label{GRt}
t - t_0 = \int G(V)^{-1/2} d V,
\end{equation}
where function $G(V)$ is defined by
\begin{equation}
G(V) \equiv 3\Lambda V^2 + 3 k_4^2 \rho_0 V^{2-\gamma} + C.
\end{equation}
Here $C$ are constants of integration. Based on this result, the
other variables can be calculated straightforwardly and the
answers are
\begin{eqnarray}
H &=& \frac{G(V)^{1/2}}{3V},  \label{H11} \\
a_i &=& a_{0i} V^{1/3} \exp \left[ K_i \int V^{-1} G(V)^{-1/2} d V
\right], \quad i=1,2,3, \\
A &=& 3 K^2 G(V)^{-1}, \\
q &=& \frac{18\gamma k_4^2 \rho_0 V^{2-\gamma} + 12C}{4G(V)} - 1.
\end{eqnarray}
Here the $a_{0i}, \, K_i, \, i=1,2,3$ are constants of
integration and $K^2=\sum_{i=1}^3 K_i^2$. In addition, the
arbitrary integration constants $K_i$ and $C$ must satisfy the
consistency condition
\begin{equation}
K^2=\frac23 C.
\end{equation}
Although the general solution can only be expressed in the
parametric form via the volume scale factor $V$, the behaviors of
the physical variables, namely the mean anisotropy parameter $A$
and the deceleration parameter $q$, can be plotted with respect to
the time as shown in the Fig.\ref{fig:F1} and Fig.\ref{fig:F2}.

Moreover, from the the above parametric expressions, we can still
analyze the physics in two different interesting limits. First of
all, let us consider the large $t$ (or equivalently large $V$)
limit. Since the value of parameter $\gamma$ is in the range $1
\le \gamma \le2$, the asymptotic value of $G(V)$ will approach
$G(V) \to  3 \Lambda V^2$ when the cosmological constant is
non-vanishing ($\Lambda
> 0$). This leads the volume scale parameter of our Universe to
expand exponentially in the large time limit, i.e. $V \propto
\exp[ \sqrt{3 \Lambda} t ]$. Therefore, in this limit, the mean
anisotropy parameter decays to zero exponentially, $A \propto \exp
[ -2 \sqrt{3 \Lambda} t ] \to 0$, and the deceleration parameter
becomes negative $q \propto \exp [ -\gamma \sqrt{3 \Lambda} t ]  -
1 < 0$. Hence the Universe can be isotropized dynamically and
undergoes an accelerated expansion in the large time limit due to
the presence of a positive cosmological constant $\Lambda$. In
fact, the value of the cosmological constant can only change the
expanding rate of the Universe but will not affect its effect of
isotropization in general.

In order to illustrate this point, let us consider the model with
a vanishing cosmological constant $\Lambda=0$. Therefore, we have
$G(V) \propto V^{2-\gamma}$ in the large time limit. This leads to
the results $V \propto t^{2/\gamma}$, $A \propto t^{2-4/\gamma}$,
and $q \to 3\gamma/2 - 1
> 0$. Hence, the Universe can still be isotropized dynamically
except for the case when $\gamma=2$. Note that the evolution will,
however, be deaccelerated in the case $\gamma=2$.

Next, let us focuss on the earlier stages of the above exact
solutions. For simplicity, we will assume $C=0$ in this case. One
can show that $G(V) \sim V^{2-\gamma}$. Therefore, one can solve
for $V \propto t^{2/\gamma}$ from the Eq. (\ref{H11}). Hence one
has $A \propto t^{2-4/\gamma}$, and $q \to 3\gamma/2 - 1
> 0$ from the rest of the field equations. Therefore, at the
earlier period stage of the Universe, the evolution is
decelerating even a cosmological constant is present. This is also
shown numerically in the Fig.\ref{fig:F1} and Fig.\ref{fig:F2}.
Moreover, the mean anisotropy parameter $A$ is always
non-vanishing independent of the values of $\gamma$. This means
that the early Universe is always anisotropic. Therefore, for the
Universe with perfect fluid matter in the CET, the initial state
is always anisotropic and this primordial anisotropy is smeared
away as a consequence of the evolution of the Universe.

\subsection{Brane Cosmology}
We will focuss on the brane effect for the model with a perfect
fluid in this subsection.  The same constraint for the perfect
fluid conservation law (\ref{GRrho}) still holds on the brane. In
addition, the gravitational field equations and the conservation
law on the brane take the form, in terms of the variables
(\ref{V})-(\ref{H}),
\begin{eqnarray}
3 \dot H + \sum_{i=1}^3 H_i^2 &=& \Lambda - \frac{3\gamma-2}2 \,
k_4^2 \rho - \frac{3\gamma-1}{12} \, k_5^4 \rho^2, \label{BWH} \\
\frac1{V} \frac{d}{d t} (V H_i) &=& \Lambda - \frac{\gamma-2}2 \,
k_4^2 \rho - \frac{\gamma-1}{12} \, k_5^4 \rho^2, \quad i=1,2,3,
\label{BWHi} \\
\dot \rho + 3 \gamma H \rho &=& 0. \label{BWrho}
\end{eqnarray}
By comparing the above equations with the ones we considered in
the previous section for the CET, it is easy to realize that the
difference is the quadratic effect due to the energy density
$\rho$. Moreover, the brane cosmology will reduce to CET if we
take the limit $k_5 \to 0$ and adjust the value of $k_4$
accordingly. We will perform the stability analysis, in the next
section, for the brane universe. The CET can be reproduced by
imposing the limit $k_5 \to 0$.

The general solution of the above system was obtained in the exact
parametric form in \cite{CHM01} by the same approach used in the
previous section. Instead of expressing results as functions of
time, we are able to present all variables, including time, in
terms of volume scale factor, $V$, with $V \geq 0$. For instance,
the time variable can be expressed as
\begin{equation}\label{tt0}
t - t_0 = \int F(V)^{-1/2} d V,
\end{equation}
where $F(V)$ is defined as
\begin{equation}
F(V) \equiv 3\Lambda V^2 + 3 k_4^2 \rho_0 V^{2-\gamma} + \frac14
k_5^4 \rho_0^2 V^{2-2\gamma} + C, \label{F11}
\end{equation}
with $\rho_0$ and $C$ the constants of integration. The other
variables are
\begin{eqnarray}
H &=& \frac{F(V)^{1/2}}{3V}, \\
a_i &=& a_{0i} V^{1/3} \exp \left[ K_i \int V^{-1} F(V)^{-1/2} d V
\right], \quad i=1,2,3, \\
A &=& 3 K^2 F(V)^{-1}, \\
q &=& \frac{3\gamma(6 k_4^2 \rho_0 V^{2-\gamma} + k_5^4 \rho_0^2
V^{2-2\gamma}) + 12C}{4F(V)} - 1,
\end{eqnarray}
where $a_{0i}, \, K_i, \, i=1,2,3$ are constants of integration
and $K^2=\sum_{i=1}^3 K_i^2$. In addition, the arbitrary
integration constants $K_i$ and $C$ must satisfy the same
consistency condition $K^2=2C/3$.

\begin{figure}
\includegraphics[width=8cm]{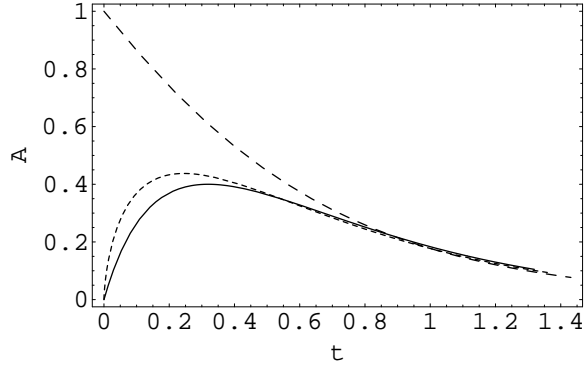}
\caption{\label{fig:F3} Mean anisotropy parameter of the Bianchi
type I brane Universe with a perfect fluid: $\gamma=2$ (solid
curve), $\gamma=1.5$ (dotted curve) and $\gamma=1$ (dashed curve).
The normalization of the parameters is chosen as $3\Lambda=1,
3k_4^2\rho_0=2, k_5^4\rho_0^2=4$, and $C=1$.}
\end{figure}

One can immediately show that the effect of the energy density
quadratic term becomes significant at the high energy epoch, or in
another words, at the early stages of the Universe by looking at
the Eq. (\ref{F11}). Indeed, $F(V) \propto V^{2-2 \gamma}$ at the
limit $t \to 0$ when $V$ is extremely small. As a result, the $F$
diverges as $t \to 0$. Therefore, the mean anisotropy parameter $A
\to 0$ at the early universe. On the other hand, in the large time
limit, the properties of the Universe should be more or less the
same as the case we have discussed for the CET in the previous
section by looking at the same Eq. (\ref{F11}). Hence the early
evolution of the anisotropic Bianchi type I brane Universe is
dramatically changed due to the presence of the brane correction
terms proportional to the square of the energy density. The time
variation of the mean anisotropy parameter of the Bianchi type I
space-time is presented, for different values of $\gamma$, in
Fig.\ref{fig:F3}. From the Fig.\ref{fig:F3}, it is clear that at
high energy density the evolution of the brane Universe always
starts out from an isotropic state with $A \to 0$. The mean
anisotropy parameter increases and reaches a maximum value after a
finite time interval $t_c$. One can show that, when $t>t_c$, the
mean anisotropy parameter is a monotonically decreasing function
approaching zero in the large time limit. This behavior is in
sharp contrast to the usual evolution in the CET, as shown in
Fig.\ref{fig:F1}, in which the Universe has to kick off from state
of maximum anisotropy due to the constraint from the field
equation.

In addition, the early time evolution of the brane universe is
normally not in an inflationary phase. On the other hand,  the
brane Universe always ends up in an inflationary phase in the
large time limit in the presence of a nonvanishing cosmological
constant. These are generic features of the brane Universe due to
the constraint of the field equations on the brane cosmology.
Indeed, a more detailed information can be extracted from the Eq.
(\ref{tt0}) in the limit $t \to 0$, or equivalently, the case with
a small $V$. For simplicity, one will take $C=0$ again. Indeed, we
can show that $V \propto t^{1/\gamma}$ as $t \to 0$. Hence, the
expansion of the early universe is of the form of power law
expansion. In addition, in the early stages of evolution of the
brane Universe the mean anisotropy parameter varies as $A \propto
t^{2-2/\gamma}$ approaching zero as $t \sim 0$.
Moreover, the deceleration parameter is given by $q=3\gamma-1$
which is always positive for all possible values of $\gamma$ for
the case $C=0$.

\begin{figure}
\includegraphics[width=8cm]{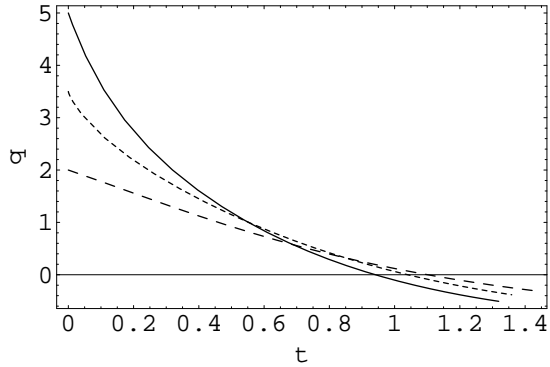}
\caption{\label{fig:F4} Deceleration parameter of Bianchi type I
brane universe with a perfect fluid: $\gamma=2$ (solid curve),
$\gamma=1.5$ (dotted curve) and $\gamma=1$ (dashed curve). The
normalization of the parameters is chosen as $3\Lambda=1,
3k_4^2\rho_0=2, k_5^4\rho_0^2=4$, and $C=1$.}
\end{figure}

\subsection{Stability Analysis}

The general perturbations for the FRW background with perfect
fluid can be found in Ref. \cite{LMSW01}. The same consideration
is, however, more complicate for the anisotropic background. For
the primary effect, we will only consider the scalar mode and
neglect the vector and tensor modes \cite{CK01,Ko01}. The metric
perturbation considered here is
\begin{equation}
a_i \to a_{Bi} + \delta a_i = a_{Bi} (1 + \delta b_i),
\end{equation}
and the perturbations with respect to the perfect fluid considered
in this paper is
\begin{equation}\label{rhoPer}
\rho \to \rho_B + \delta \rho, \qquad p \to p_B + (\gamma - 1)
\delta \rho.
\end{equation}
Here the variables with subscript $B$ are the exact solutions
presented in the previous sections. For technical convenience, we
will use variables $\delta b_i$ instead of $\delta a_i$ in our
analysis. Therefore, the perturbations of the following quantities
can be shown to be
\begin{equation}\label{QPer}
H_i \to H_{Bi} + \delta\dot b_i, \quad H \to H_B + \frac13 \sum_i
\delta\dot b_i, \quad \sum_i H_i^2 \to \sum_i H_{Bi}^2 + 2 \sum_i
H_{Bi}\, \delta\dot b_i, \quad V \to V_B + V_B \sum_i \delta b_i.
\end{equation}
As discussed in the previous sections, the results in the CET are
special cases for the brane universe, in the limit $k_5=0$.
Therefore, we can perform our stability analysis only for the
brane universe first. The conclusion for the CET can be easily
extracted from the brane world case.

The perturbation equations for various quantities, can be obtained
by  substituting the perturbations (\ref{rhoPer}, \ref{QPer}) into
the field equations (\ref{BWH}, \ref{BWHi}, \ref{BWrho}). Leading
terms will reproduce the field equations.  Therefore, one has the
following perturbation equations from the first order terms
$O(\delta b_i,  \, \delta \rho)$,
\begin{eqnarray}
\sum_{i=1}^3 \delta \ddot b_i + 2 \sum_{i=1}^3 H_{Bi} \, \delta
\dot b_i &=& - \frac{3\gamma-2}2 \, k_4^2 \, \delta \rho -
\frac{3\gamma-1}6 \, k_5^4 \rho_B \, \delta \rho, \label{HPer} \\
\delta \ddot b_i + \frac{\dot V_B}{V_B} \delta \dot b_i + H_{Bi}
\sum_{j=1}^3 \delta \dot b_j &=& - \frac{\gamma-2}2 \, k_4^2 \,
\delta \rho - \frac{\gamma-1}6 \, k_5^4 \rho_B \, \delta \rho,
\quad i=1,2,3, \label{HiPer} \\
\delta \dot \rho + 3 \gamma H_B \, \delta \rho + \gamma
\sum_{i=1}^3 \delta \dot b_i \, \rho_B &=& 0.
\end{eqnarray}

In order to solve the above system of differential equations, we
need an inspiration from the process of constructing the exact
solutions. First of all, from the general results (\ref{rho},
\ref{Hi}) in both the CET and the brane theory, their dynamics on
the perturbation variables follow the constraints
\begin{eqnarray}
\delta \rho &=& - \gamma \rho_B \sum_{i=1}^3 \delta b_i,
\label{P1} \\
\delta \dot b_i &=& \frac13 \sum_{j=1}^3 \delta \dot b_j -
\frac{K_i}{V_B} \sum_{j=1}^3 \dot b_j. \label{P2}
\end{eqnarray}
By summing the Eqs. (\ref{HiPer}) and then using the result
(\ref{P1}), we end up with a second order differential equation
for a variable of the combination $\sum_{i=1}^3 \delta b_i$
\begin{equation} \label{EqPer}
\sum_{i=1}^3 \delta \ddot b_i + 6 H_B \sum_{i=1}^3 \delta \dot
b_i - \frac12 \gamma \rho_B [ 3(\gamma-2) k_4^2 + (\gamma-1)
k_5^4 \rho_B ] \sum_{i=1}^3 \delta b_i = 0.
\end{equation}
The task is to solve ${\sum \delta b_i}$ from the above equation
and then to construct $\delta b_i$ and $\delta \rho$ from the Eqs.
(\ref{P2}) and (\ref{P1}). For the purpose of the stability
analysis in the final stage, it is sufficient to consider the
large time limit behaviors of the perturbation variables $\delta
b_i$ and $\delta \rho$. From the discussion on the asymptotic
behavior of the exact solutions, qualitative outcome in both the
CET and the brane theory, we should divide our analysis into two
different cases in the presence of absence of the cosmological
constant.

For the case with a positive cosmological constant, one can
extract the asymptotic forms of the background variables from the
exact solutions which give
\begin{equation}
H_B \to \sqrt{\Lambda/3}, \qquad V_B \propto \exp(\sqrt{3\Lambda}
\, t), \qquad \rho_B \propto \exp(-\sqrt{3\Lambda} \, \gamma t).
\end{equation}
The asymptotic expression of $\rho_B$ indicates that the third
term in Eq. (\ref{EqPer}) can be neglected in the large time
limit. As a result, we have the following equation for the
asymptotic $\sum_{i=1}^3 \delta b_i$ which remains valid in both
the CET and the brane world
\begin{equation}
\sum_{i=1}^3 \delta \ddot b_i + \sqrt{12\Lambda} \sum_{i=1}^3
\delta \dot b_i = 0.
\end{equation}
This in term leads to the final result
\begin{equation}
\sum_{i=1}^3 \delta b_i \propto \exp(-\sqrt{12\Lambda}\, t).
\end{equation}
Therefore, from Eqs. (\ref{P1}, \ref{P2}) we can obtain the
asymptotic expressions of the following perturbation variables
\begin{equation}
\delta b_i \propto \exp(-\sqrt{12\Lambda}\, t), \qquad \delta \rho
\propto \exp[ -\sqrt{3\Lambda}\,(\gamma+2) t].
\end{equation}
This indicates that that the background solutions are stable.

Similarly, for the case with $\Lambda=0$, we have
\begin{equation}
H_B \to \frac2{3\gamma} t^{-1}, \qquad V_B \to \left(
\frac{\sqrt{3 \rho_0 k_4^2}\, \gamma}2 \, t \right)^{2/\gamma},
\qquad \rho_B \to \frac4{3 k_4^2 \gamma^2} \, t^{-2},
\end{equation}
where the preceding coefficients become important in this case.
Therefore, the Eq. (\ref{EqPer}), when the brane correction due to
the quadratic energy density term signified by the parameter $k_5$
can be neglected in the large time limit, reduces to
\begin{equation} \label{EqPer1}
\sum_{i=1}^3 \delta \ddot b_i + \frac4{\gamma} t^{-1} \sum_{i=1}^3
\delta \dot b_i - \frac{2(\gamma-2)}{\gamma} t^{-2} \sum_{i=1}^3
\delta b_i = 0.
\end{equation}
The solution of the above equation has the form
\begin{equation}
\sum_{i=1}^3 \delta b_i \propto t^\alpha,
\end{equation}
where the exponent parameter is determined by
\begin{equation}
\alpha (\alpha - 1) + \frac4{\gamma} \alpha -
\frac{2(\gamma-2)}{\gamma} = 0.
\end{equation}
The explicit expression of $\alpha$ is
\begin{equation}
\alpha = \frac{(\gamma-4) \pm \sqrt{(4-\gamma)^2 +
8\gamma(\gamma-2)}}{2\gamma} \le 0,
\end{equation}
which is always negative for all possible values of $\gamma$. The
only exception is the case when $\gamma=2$ such that $\alpha=0$ is
a possible solution. In summary, the asymptotic behavior of
perturbation variables $\delta b_i$ and $\delta \rho$ is
\begin{equation}
\delta b_i \propto t^\alpha, \qquad \delta \rho \propto
t^{\alpha-2},
\end{equation}
which shows that the background solutions are always stable even
when the cosmological constant is vanishing.

\section{Anisotropic Universe with Scalar Field}
In the previous section we have considered the evolution of the
universe with a perfect fluid, obeying a barotropic equation of
state, in the anisotropic background. We will study the stability
problem of the anisotropic universe with a scalar field in this
section. It is known that the scalar field $\phi$ with a potential
can be thought of as a perfect fluid with the energy density and
pressure given by $\rho_\phi = \dot\phi^2/2 + U(\phi)$ and $p_\phi
= \dot\phi^2/2 - U(\phi)$ respectively. Here $U(\phi)$ is the
scalar field potential. The scalar field is expected to play a
fundamental role in the evolution of the early Universe.

In this section, we will only consider the case where the scalar
field potential is a constant, $U(\phi)=\Lambda=\text{constant}>0$
acting as a cosmological constant. As discussed earlier, the
related anisotropic cosmology in the CET is just a particular
limit of the brane universe. Therefore, we will consider the brane
universe in details for the moment. The CET  can be recovered by
taking a suitable limit.

When we discuss the stability problem of the system, we will
focuss on the system with a scalar potential that admits at least
a local minimum $\phi_0$ such that $U(\phi=\phi_0)=\Lambda$. We
will assume that $\phi_0$ is the asymptotic solution to the field
equation in the large time limit. We will able to show that the
system remains stable in the large time limit even the
evolutionary solution is only known when the scalar potential is a
constant.

\subsection{Brane Cosmology}

For a Bianchi type I brane Universe with a scalar field the
gravitational field equations take the following form
\cite{CHM01}:
\begin{eqnarray}
3 \dot H + \sum_{i=1}^3 H_i^2 &=& k_4^2 \left[ U(\phi) -
\dot\phi^2 \right] - k_5^4 \left[ \frac16 \left(
\frac{\dot\phi^2}2 + U(\phi) \right)^2 + \frac14 \left(
\frac{\dot\phi^4}4 - U^2(\phi) \right) \right], \label{SdH} \\
\frac1{V} \frac{d}{d t}(VH_i) &=& k_4^2 U(\phi) - \frac1{12} k_5^4
\left( \frac{\dot\phi^4}4 - U^2(\phi) \right), \quad i=1,2,3.
\label{SdV}
\end{eqnarray}
Note that the scalar field also obeys the following evolution
equation:
\begin{equation}\label{ddphi}
\ddot \phi + 3 H \dot\phi + \frac{dU(\phi)}{d\phi} = 0.
\end{equation}
By imposing the constraint $U(\phi)=\Lambda$, the Eq.
(\ref{ddphi}) gives $\dot\phi=2\phi_0/V$ with $\phi_0>0$ a
constant of integration. Summing the Eq. (\ref{SdV}) over the
index $i$ and comparing with the Eq. (\ref{SdH}), one can extract
the following equation:
\begin{equation}\label{SddV}
\ddot V = \kappa^2 V - \kappa_0^2 V^{-3}.
\end{equation}
Here we have denoted the positive constants $\kappa$ and
$\kappa_0$ as $\kappa^2 = 3k_4^2\Lambda + k_5^4\Lambda^2/4$ and
$\kappa_0^2 = k_5^4\phi_0^4$.

The general solution of the Eq. (\ref{SddV}) can be shown to be
\cite{CHM01}
\begin{equation}
V(t) = \frac1{2\kappa} e^{-\kappa(t-t_0)} \sqrt{ \left[
e^{2\kappa(t-t_0)} - C \right]^2 - 4 \kappa_0^2 \kappa^2}.
\end{equation}
Here $C$ is an integration constant.

The time evolution of the expansion, scale factors, mean
anisotropy, shear and deceleration parameter are given by
\begin{eqnarray}
H(t) &=& \frac{\kappa}3 \frac{e^{4\kappa(t-t_0)} - C^2 + 4
\kappa_0^2 \kappa^2}{ \left( e^{2\kappa(t-t_0)} - C \right)^2 - 4
\kappa_0^2 \kappa^2}, \label{2H} \\
a_i(t) &=& a_{0i} e^{-\kappa(t-t_0)/3} \left[ \left(
e^{2\kappa(t-t_0)} - C \right)^2 - 4 \kappa_0^2 \kappa^2
\right]^{1/6} \exp \left[ 2K_i F \left( e^{\kappa(t-t_0)} \right)
\right], \quad i=1,2,3, \\
A(t) &=& \frac{12K^2 e^{2\kappa(t-t_0)} \left[ \left(
e^{2\kappa(t-t_0)} - C \right)^2 - 4 \kappa_0^2 \kappa^2 \right]}{
\left[ e^{4\kappa(t-t_0)} - C^2 + 4 \kappa_0^2 \kappa^2
\right]^2}, \\
q(t) &=& 12 e^{2\kappa(t-t_0)} \frac{ C e^{4\kappa(t-t_0)} + (C^2
- 4 \kappa_0^2 \kappa^2) \left[ C - 2 e^{2\kappa(t-t_0)} \right]}{
\left[ e^{4\kappa(t-t_0)} - C^2 + 4 \kappa_0^2 \kappa^2
\right]^2} - 1, \label{2q}
\end{eqnarray}
where $a_{0i}, \, i=1,2,3$ are arbitrary constants of integration
and $F(x)=\int [ (x^2-C)^2 - 4\kappa_0^2\kappa^2 ]^{-1/2} dx$. We
do expect that the Universe starts its evolution from a singular
initial condition. Therefore, according to the value $V(0)$, the
parameter $t_0$ should be chosen as $\exp(-2\kappa
t_0)=C+2\kappa_0\kappa$.

Note that we can analyze the evolution of the Universe in two
different regions. First of all, at the very early time, the
parameters can be shown to be
\begin{equation}
V \sim \sqrt{2\kappa_0 t} \sim 0, \qquad A \sim 6 K^2 t/\kappa_0
\sim 0, \qquad q \sim 5 > 0.
\end{equation}
Secondly, the asymptotic behavior of the parameters can be shown
to be
\begin{equation}
V \propto e^{\kappa t}, \qquad A \propto e^{-2\kappa t}, \qquad q
\to -1 < 0.
\end{equation}
This result is similar to the model with a perfect fluid such that
the brane Universe both evolves from an isotropic singular state
initially. Later on, the mean anisotropy parameter $A$ increases
dynamically and decays to zero after the mean anisotropy parameter
reaches its maximal value.

\subsection{Conventional Einstein's Theory}

The conventional Einstein's theory (CET)  can be easily deduced
from the above results by taking the limit $k_5=0$, or, in this
case, $\kappa_0=0$. The result is
\begin{eqnarray}
V(t) &=& \frac{\sqrt{C}}{2\kappa} \left( e^{\kappa t} - e^{-\kappa
t} \right), \\
H(t) &=& \frac{\kappa}3 \frac{e^{2\kappa t} + 1}{ e^{2\kappa t} -
1}, \\
a_i(t) &=& a_{0i} C^{1/6} \left( e^{\kappa t} - e^{-\kappa t}
\right)^{1/3} \exp \left[ 2K_i C^{-1} \int ( e^{2\kappa x} -
1)^{-1} d x \right], \quad i=1,2,3, \\
A(t) &=& 12K^2 C^{-1} \left( e^{\kappa t} + e^{-\kappa t}
\right)^{-2} \\
q(t) &=& 12 \left( e^{\kappa t} + e^{-\kappa t} \right)^{-2}-1.
\end{eqnarray}
Note that $\exp(-2\kappa t_0)=C$. The asymptotic behavior is the
same as the case in the brane world. The initial state is,
however, very different such that
\begin{equation}
V \sim \sqrt{C} t \sim 0, \qquad A \sim 3 K^2/C \sim
\hbox{constant}, \qquad q \sim 2 > 0.
\end{equation}
Therefore, for the CET, the Universe has to start out from an
anisotropic initial expansion and then decays all the way to the
phase of isotropic expansion in the large time limit.

\begin{figure}
\includegraphics[width=8cm]{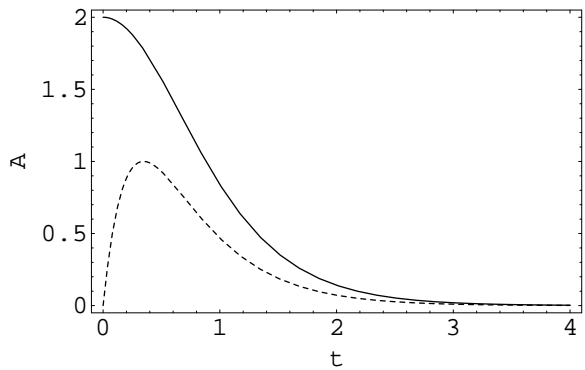}
\caption{\label{fig:F5} Mean anisotropy parameter for the  Bianchi
type I universe with a scalar field: for the CET (solid curve) and
for the brane world (dotted curve). The normalization of the
parameters is set as $\kappa=1, 2\kappa_0\kappa=C$.}
\end{figure}

The evolution of the mean anisotropy parameter $A(t)$ for the CET
and the brane world is presented in the Fig.\ref{fig:F5}.

\begin{figure}
\includegraphics[width=8cm]{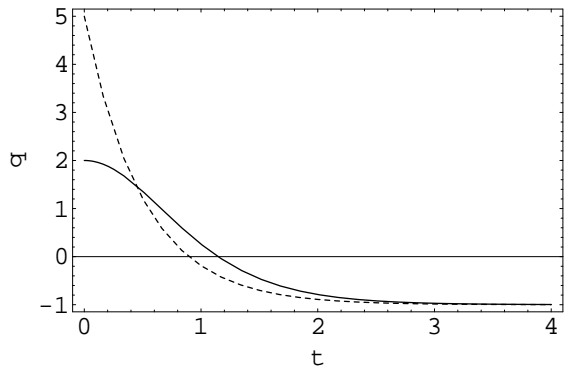}
\caption{\label{fig:F6} Deceleration parameter for the  Bianchi
type I universe with a scalar field: for the CET (solid curve) and
for the brane world (dotted curve). The normalization of the
parameters is set as $\kappa=1, 2\kappa_0\kappa=C$.}
\end{figure}

\subsection{Stability Analysis}

Since the the quadratic brane correction plays a significant role
only at the early stage of the evolution of the Universe. For the
purpose of our stability analysis in the large time limit, we will
focuss on the asymptotic behavior of the fields. Therefore, it is
sufficient to consider the background solutions in the
conventional Einstein's theory.

Perturbations of the fields of a gravitational system against the
background evolutionary solution should be checked to ensure the
stability of the exact or approximated background solution. In
principle, the stability analysis should be performed against the
perturbations of all possible fields in all possible manners
subject to the field equations and boundary conditions of the
system. In the following section, we will divide the perturbations
into two disjoint classes: (a) the perturbations of the scale
factors, or equivalently the metric field; and (b) the
perturbations of the dilaton field.

We will argue that the most complete stability conditions we are
looking for can be obtained from class (a) and class (b)
perturbations; even the backreaction of the scalar field
perturbation on the metric field perturbations is known to be
important \cite{back}. We will show that this backreaction does
not bring in any further restriction on the stability conditions.

The reason is rather straightforward. One can write the linearized
perturbation equation as
\begin{equation}
D^i_{a_j} \delta a_j + D^i_\phi \delta \phi = 0,
\end{equation}
for the system we are interested. Moreover, perturbations are
defined as $a_i = a_i^0 + \delta a_i$ and $\phi = \phi_0 + \delta
\phi$ with the index $0$ denoting the background field solution.
Note also that the operators $D^i_{a_j}$ and $D^i_\phi$ denote
the differential operator one obtained from the linearized
perturbation equation with all fields evaluated at the background
solutions. The exact form of these differential operators will be
shown later in the following arguments.

One is looking for stability conditions that the field parameters
must obey in order to keep the evolutionary solution stable. One
can show that class (a) and class (b) solutions are good enough to
cover all domain of stability conditions. Let us denote the domain
of solutions to class (a), (b), and (a+b) stability conditions as
$S(a), S(b)$, and $S(a+b)$ respectively. Specifically, the
definition of these domains are defined by $S(a) \equiv \{ \delta
a_i | D^i_{a_j} \delta a_j = 0 \}$, $S(b) \equiv \{ \delta \phi |
D^i_\phi \delta \phi = 0 \}$, and $S(a+b) \equiv \{ ( \delta a_i,
\delta \phi ) | D^i_{a_j} \delta a_j + D^i_\phi \delta \phi = 0
\}$.

Therefore, one only needs to show that $S(a) \cap S(b) \subset
S(a+b)$. This is because that ``$D^i_{a_j} \delta a_j = 0$ and
$D^i_\phi \delta \phi = 0$'' imply that ``$D^i_{a_j} \delta a_j +
D^i_\phi \delta \phi = 0$.'' On the other hand, ``$D^i_{a_j}
\delta a_j + D^i_\phi \delta \phi = 0$'' does not imply that
``$D^i_{a_j} \delta a_j =0$ or $D^i_\phi \delta \phi =0$.'' Hence
class (a) and class (b) solutions cover all the required
stability conditions we are looking for. Hence we only need to
consider these two separate cases for simplicity.

In addition, one knows that any small time-dependent perturbation
against the metric field is known to be equivalent to a gauge
choice \cite{gauge}. This can be clarified as follows. Indeed, one
can show that any small coordinate change of the form $x'^\mu =
x^\mu - \epsilon^\mu$ will induce a gauge transformation on the
metric field according to $g'_{\mu\nu}=g_{\mu\nu} + D_\mu
\epsilon_\nu + D_\nu \epsilon_\mu$. Therefore, a small metric
perturbation against a background metric is amount to a gauge
transformation of the form $a_i' = a_i + \epsilon^t \dot a_i$ for
the Bianchi type-I metric with $\epsilon_\mu =({\rm constant},
\epsilon_i(t))$. This is then equivalent to small metric
perturbations. If a background solution is stable against small
perturbation with respect to small field perturbations, one in
fact did nothing but a field redefinition.

If the background solution is, however, unstable against small
perturbations, e.g., the small perturbation will grow
exponentially as we will show momentarily, the resulting large
perturbations cannot be classified as small gauge transformation
any more. Therefore, the stability analysis performed in the
literature \cite{jb1,jb2,kp91,dm95,dm2,KPZ99,kao00,kim,abel} for
various models against the unstable background solution served as
a very simple method to check if the system supports a stable
metric field background. This is the reason why we still perform a
perturbation on the metric field for stability analysis; even a
small perturbation is equivalent to a gauge redefinition.

Note that one should also consider a more general perturbation
with space perturbation included. The formulation is, however,
much more complicated than the one we will show in this paper. We
will focus on the time-dependent case for simplicity in this
paper. The space-dependent perturbation analysis is still under
investigation. The time-dependent analysis alone will, however,
bring us much useful information for the stability conditions
about the model we are interested. For example, we will show in
the following subsection that the solution found in Ref.
\cite{CHM01} remains stable as long as the scalar field falls
close to any local minimum of the potential $U(\phi)$. Note again
that the solution found in Ref. \cite{CHM01} is an exact solution
only when $U=$ constant.
\bigskip

\centerline{\bf Dilaton Perturbation}

Let us consider the perturbation of the dilaton field of the
following form
\begin{equation}
\phi \to \phi_B + \delta\phi, \qquad U(\phi) \to \Lambda +
\partial_\phi \delta\phi.
\end{equation}
By setting $k_5=0$ in the Eqs. (\ref{SdH}, \ref{SdV},
\ref{ddphi}), the field equations for $\delta\phi$ become
\begin{eqnarray}
-2 \dot \phi_B \delta \dot \phi + \partial_\phi U \delta \phi
&=& 0, \\
\partial_\phi U \delta \phi &=& 0, \\
\delta \ddot \phi + 3 H_B \delta \dot \phi + \partial_\phi^2 U
\delta \phi &=& 0.
\end{eqnarray}
Due to the fact that $H_{Bi} \to \frac{\kappa}3$, the asymptotic
solution of $\delta \phi$ can be shown to be
\begin{equation}
\delta\phi \propto \exp\left[ (-\kappa \pm \sqrt{\kappa^2-4\beta})
t / 2 \right].
\end{equation}
Here $\beta \equiv \partial_\phi^2U|_{t\to\infty}$. Therefore, the
perturbation of dilaton field decays to zero if $\beta\ge 0$.
Hence we show that the system is stable with respect to the scalar
perturbation if $\beta\ge 0$ such that the scalar potential is
capable of confining the scalar field to its local minimum.

Note that there are two additional equation $U'=0$ and
$\dot{\phi}=0$ which is required for consistency of the stability
of the system. These constraints simply imply that the scalar
field must be at rest at the local minimum of the scalar field
potential. As we have pointed out earlier that the background
solution we have at hand is an exact solution for the model with a
constant cosmological constant. Therefore, the background solution
we used for stability analysis is only an approximated solution
which remains valid only when the scalar field is close to the
local minimum of the the scalar field potential. As a result, one
should not be too serious about these two further constraints. In
fact, these constraints are both negligible in the large time
limit when the scalar field falls close to the local minimum of
the scalar field potential, namely, $U' \to 0$ and $\dot{\phi} \to
0$. Hence one is able to show that these additional constraints
can be made satisfied approximately in the large time limit.



\bigskip

\centerline{\bf Metric Perturbation}

Using the metric perturbation (\ref{QPer}), the perturbation
equations for the metric perturbation $\delta b_i$ can be obtained
from perturbing the Eqs. (\ref{SdH}, \ref{SdV}, \ref{ddphi}). The
result is
\begin{eqnarray}
\sum_i \delta \ddot b_i + 2 \sum_i H_{Bi} \delta \dot b_i &=& 0,\\
\delta \ddot b_i + \frac{\dot V_B}{V_B} \delta \dot b_i + H_{Bi}
\sum_j \delta \dot b_j &=& 0,\\
\dot \phi_B \sum_i \delta \dot b_i &=& 0.
\end{eqnarray}
Here we also choose the limit $k_5=0$. Note that the background
variables $V_B, a_{Bi}$ and $H_{Bi}$ approach
\begin{equation}
V_B \propto e^{\kappa t}, \qquad a_{Bi} \propto e^{\kappa t/3},
\qquad H_{Bi} \to \frac{\kappa}3.
\end{equation}
Therefore, the asymptotic behavior of the metric perturbation
$\delta b_i$ can be found from the second equation. In addition,
the other two field equations will provide a constraint for the
system. The results are
\begin{equation}
\delta b_i \to c_i e^{-\kappa t}, \qquad \sum_i c_i = 0.
\end{equation}
Hence, the metric perturbation
\begin{equation}
\delta a_i \equiv a_{Bi} \delta b_i \to e^{-2\kappa t/3} \to 0,
\end{equation}
in the large time limit. This indicates that the background
solution of the system is stable against the metric perturbation
as shown above.

\section{Conclusion}
We have discussed the anisotropic property of cosmological models
in the CET and brane theory. A realistic model, being consistent
with the current observations, should produce a small value of
anisotropic parameter at the later stage of the evolution of our
Universe near the last scattering surface. By assuming the Bianchi
type I space-time for the evolution of our Universe, we found that
the final state of the evolving Universe always approaches the
phase of isotropic expansion in both theories.

These two different theories give completely different initial
anisotropy at the very early stage of evolution.  Indeed, for the
CET, the anisotropy tends to be large in the very early stage. In
another words, the universe tends to begin from a highly
anisotropic initial state. The mean anisotropy parameter $A$ will
then decay to zero as the time increases. On the other hand, the
early time behavior of the Universe in the brane world scenario
changes significantly due to the quadratic correction on the
brane. As a result, any non-vanishing mean anisotropy parameter,
$A(t)$, tends to vanish in the very early period. There is a
characteristic time, $t_c$, that divides the evolution of $A(t)$
into two different stages. The mean anisotropy parameter is
increasing when $t<t_c$ and reaches its maximal value at $t=t_c$.
After that, $A(t)$ starts to decay. This kind of behavior is
clearly shown in the Fig.\ref{fig:F2} and Fig.\ref{fig:F5}. This
result remains true for both the model with a perfect fluid and
the model with a scalar field. And this appears to be a general
feature independent of the types of matter considered.

It is worth noting that the only exception is the model with $p=0$
(i.e. $\gamma=1$) of the perfect fluid model. The mean anisotropy
of this model behaves similar to the models in the CET where mean
anisotropy parameter is large in the very early time. Moreover, we
also analyzed the stability problem for those exactly solved
anisotropic models shown in this paper. The result indicates that
all of the solution known to us  are stable in the large time
limit. Therefore, the evolution of the Universe in the CET starts
with highly initial anisotropic expansion. The dynamic of the
system will take the Universe to the phase of isotropic expansion
in the large time limit. We also show that the final isotropic
expansion will remain stable in the large time limit. In addition,
the mean anisotropy parameter will keep decreasing as time
increases. The model provided here is a useful and explicit model
that is capable of providing us with a Universe that has a tiny
anisotropy left over near the last scattering surface.

\section*{Acknowledgments}
This work is supported in part by the National Science Council
under the grant numbers NSC90-2112-M009-021 and
NSC90-2112-M002-055. The work of CMC is also supported by the
Taiwan CosPA project and, in part, by the Center of Theoretical
Physics at NTU and National Center for Theoretical Science. One of
the authors (WFK) is grateful to the hospitality of the physics
department of National Taiwan University where part of this work
is done.

\end{document}